# Stacked U-Nets with Self-Assisted Priors Towards Robust Correction of Rigid Motion Artifact in Brain MRI

Mohammed A. Al-masni[1], Seul Lee[1], Jaeuk Yi[1], Sewook Kim[1], Sung-Min Gho[2], Young Hun Choi[3], and Dong-Hyun Kim[1,*]

[1] Department of Electrical and Electronic Engineering, College of Engineering, Yonsei University, Seoul, Republic of Korea
[2] GE Healthcare, Seoul, Republic of Korea
[3] Department of Radiology, Seoul National University Hospital, Seoul, Republic of Korea.

**Abstract**

Magnetic Resonance Imaging (MRI) is sensitive to motion caused by patient movement due to the relatively long data acquisition time. This could cause severe degradation of image quality, and therefore affect the overall diagnosis. In this paper, we develop an efficient retrospective deep learning method called stacked U-Nets with self-assisted priors to address the problem of rigid motion artifacts in MRI. The proposed work exploits the usage of additional knowledge priors from the corrupted images themselves without the need for additional contrast data. The proposed network learns missed structural details through sharing auxiliary information from the contiguous slices of the same distorted subject. We further design a refinement stacked U-Nets that facilitates preserving of the image spatial details and hence improves the pixel-to-pixel dependency. To perform network training, simulation of MRI motion artifacts is inevitable. The proposed network is optimized by minimizing the structural similarity (SSIM) loss using the synthesized motion-corrupted images from 83 real motion-free subjects. We present an intensive analysis using various types of image priors: the proposed self-assisted priors and priors from other image contrast of the same subject. The experimental analysis proves the effectiveness and feasibility of our self-assisted priors since it does not require any further data scans. The overall image quality of the motion-corrected images via the proposed motion correction network is significantly improved in term of SSIM from 71.66% to 95.03% and the mean square error was also declined from 99.25 to 29.76. These results indicate the high similarity of the anatomical structure of the brain in the corrected images compared to the motion-free data. The motion-corrected results of both the simulated and real motion data showed the potential of the proposed motion correction network to be feasible and applicable in clinical practices.

---

[*] Corresponding author

# 1. Introduction

Magnetic Resonance Imaging (MRI) is extremely sensitive to motion caused by patient movement. This is due to the relatively long data acquisition time required to acquire the k-space data to generate the MR image [1]. Motion artifacts manifest as ghosting, ringing, and blurring and may cause severe degradation of image quality. As a result, it becomes a challenge for radiologists to accurately interpret and diagnose patients with motion artifacts, leading to increased cost if the level of artifacts is too severe and the acquisition has to be repeated [2]. Physiological motions, such as respiratory and cardiac movement, blood flow, and vessel pulsation, and voluntary and involuntary head motions, known as rigid or bulk motion, are the main sources of MRI motion artifacts [3]. Both the involuntary motions in pediatrics or neuro-degenerative patients and the conscious sudden motions due to discomfort or mental conditions are unavoidable during data acquisition. Thus, correction of motion artifacts is being more of interest for MRI society.

The type of MRI motion artifacts relies on the strategies of data acquisition, such as k-space trajectory and pulse sequence diagram, with regards to the time and amount of the subject movement [1, 4, 5]. Motion artifacts usually happened in the phase-encoding direction of the k-space, where patient motion during filling the center of k-space is relatively corresponding to the low image frequencies, resulting in motion with high severity. In contrast, patient movement during data acquisition of the edges of k-space tends to result in ringing artifacts. In general, motion artifacts reduction or correction methods can be categorized into two main groups: prospective [6, 7] and retrospective [8-11]. The former method is a real-time approach that attempts to compensate the motion artifacts during scanning by updating the pulse sequence using motion tracking devices, such as navigators or sensors. In opposite, the retrospective approach estimates the motion parameters after completing the scanning (i.e., during image reconstruction) using prior information derived from external sensors or applying iterative or autofocusing algorithms. However, such approaches require additional hardware and time, which imply a concern of calibrations with the MRI scanner and intensive computations for the estimation of motion parameters, respectively [3]. Therefore, automated correction of MRI motion artifacts without the need of external navigators is of great clinical significance and has become an active research area.

Recently, the advances of deep learning convolutional neural networks (CNNs) are gaining a lot of attention and have been widely utilized in the field of medical image analysis, specifically the MRI domain. For instance, the CNNs have been succeeded in developing several clinical MRI applications, including diagnosis of brain abnormalities [12, 13], detection and quantification of cerebral small vessel diseases [14-16], segmentation of brain tumors [17, 18], reconstruction of images [19, 20], and correction of artifacts [4, 5, 21-29]. The last few years have witnessed numerous attempts to solving the task of motion artifacts correction using the deep learning paradigm. The works in [27, 29] proposed automatic multi-stream CNNs to detect the presence of motion artifacts in cardiac MR images. A curriculum learning, a part of active learning, was employed to efficiently determine the level of motion severity by training a deep learning network with samples of gradually increasing the difficulty (i.e., from good to poor image quality) [27]. The attention supervision was used in [29] to guide the network to focus on the target region. This particular attention was achieved by computing the importance weights of the gradients with respect to the activation maps. Also, Ko et al. presented the significance of using a self-spatial attention module within the deep residual network for rigid and non-rigid motion artifacts reduction in computed tomography images [24]. In 2019, Haskell et al. developed a separable motion correction model called Network Accelerated Motion Estimation and Reduction (NAMER) [23], which integrated a deep learning motion artifacts detection with a motion estimation model. Although the NAMER has achieved promising motion mitigation, it required a long processing time of around 7.13 min for a single slice. Hence, the need for an end-to-end method with less execution time is still demanded. Similar to the NAMER, Wang et al. utilized the CNN prediction as an initial guess (i.e., data fidelity) into the optimization problem to correct the out-of-field of view motion [28]. Their results show that the combined motion model-based data fidelity outperformed the CNN prediction for the 2D motion correction. However, in the case of 3D motion correction, better image quality was obtained using the CNN prediction compared to the data fidelity model. In [5], the authors developed a densely deep residual U-Net to correct the motion artifacts in MR images. The proposed network was designed to learn the motion artifacts, where a single shortcut connection that connected the corrupted input with the residual map prediction was able to reproduce the output with reduced motion artifact. The performance of motion artifacts reduction was improved from 0.867 (the input simulated motion) to 0.965 in term of the Structural Similarity index (SSIM). Comparing to the original U-Net, their network achieved a slight improvement rate of 0.8% in term of SSIM.

Different from the above studies, [22, 25] attempted to solve the motion problem by the inclusion of multi-contrast images. Lee et al. built a framework consisting of two parts: multi-contrast image registration and motion correction networks for three MRI sequences, namely $T_1$-weighted ($T_1$w), $T_2$-weighted ($T_2$w), and Fluid-Attenuated Inversion Recovery (FLAIR) [25]. After the multi-contrast image alignment has been performed, the encoder of the motion correction network separately extracted feature vectors for each contrast image. Then, all these features were combined to share information from different contrasts. In the end, the decoder path was able to learn these features and generate multi-output corrected predictions. Even though this work presented different scenarios of using input images with motion-free or motion-corrupted, it seems that the authors did not intend to exploit multi-contrast as priors as clearly mentioned in [22]. The best motion correction scenario for any corrupted target was through feeding the rest pure contrast images without any motion. More specifically, for better motion artifact reduction in $T_1$w, it was recommended to pass the motion-free $T_2$w and FLAIR images into the network along with the corrupted $T_1$w image. Chatterjee et al. introduced a retrospective deep learning network to remove motion artifacts with the assistance of additional information presented as image priors [22]. Supplying additional prior knowledge was performed using two different types of image priors: similar slices of the image contrast from different subjects and different contrasts of the same subjects. The second prior type is similar to the work presented in [25], but it used different image contrasts, including $T_1$w, $T_2$w, and Proton Density (PD) images. During the correction process of the corrupted $T_2$w image, they utilized the motion-free $T_1$w and PD images as priors. Their findings concluded that the correction performance was not improved when the priors of similar contrasts but different subjects were utilized. In contrast, promising results were obtained in the case of using priors of different contrasts but of the same subjects.

Since it is difficult to acquire pairs of motion-corrupted images and ground-truth clean images, most research in the field of motion artifact correction trained and tested their methods using synthesized corrupted images from motion-free data. The motion artifact simulation could be achieved by applying some rotation and translation transformations in the spatial domain or by adding phase shift to the k-space in the frequency domain. The amount of motion could be controlled to derive various levels of motion severity. For example, the average SSIM for the corrupted motion data was between 40% to 74% in [28], 86.70% in [5] , and around 77.0% in [22]. However, the corrupted input images in [25] seem to be not severe since the average SSIM scores of 98.11%, 91.25%, and 92.79% were reported for T1w, T2w, and FLAIR, respectively.

It is observed that most of the aforementioned works endeavored to reduce the MRI motion artifacts through developing new deep learning architectures. However, the task of motion correction is quite different from other tasks such as segmentation and detection. It is of note that deep learning networks in segmentation and detection tasks usually try to learn some representative features from input data that reflect the target Region of Interest (ROI) (i.e., target ROI is visible to the network). Differently, deep learning motion correction networks struggle to find or correct missing structural details from the distorted images. Only a few works [22, 25] have utilized additional prior knowledge from different image contrasts of the same subjects. Nevertheless, these works required additional MRI scans, which could not be available in all routine clinical exams and does not seem to be feasible for future medical practices. Therefore, there exists sufficient room for improving the feasibility and effectivity of the motion artifacts correction.

In this paper, we address the above-mentioned issues by designing efficient stacked U-Nets with self-assisted priors to solve the problem of motion artifacts in MRI. The proposed work aims to exploit the usage of additional knowledge priors from the corrupted images themselves without the need of additional contrast data. This is achieved by sharing the structural details from the contiguous slices of the same distorted subject with each corrupted input image. More specifically, the proposed network initiates by concatenating multi-inputs (i.e., the corrupted image and its adjacent slices) and eventually yields a single corrected image. In this case, the network could reveal some missed structural details throughout the assistance of the information that exists in the adjacent slices, especially in the case of 3D imaging (as the case of this work). Furthermore, we develop stacked U-Nets that enable to re-evaluate the initial estimates by capturing some spatial relationships between predictions. Here, the prediction reuse with spatial attention preserves the spatial location and leads to better refinement of the corrected images. Finally, our proposed stacked U-Nets with the inclusion of self-assisted priors and spatial attention result in better performance of the MRI motion artifact correction.

The main contributions of this paper are outlined as follows.
1) We propose a novel motion artifact correction method that learns additional knowledge priors from the adjacent slices of the same corrupted subject. The main idea of the self-assisted priors is to capture some missed structural information from image priors and enable the network to learn these unreachable patterns to achieve more accurate correction of motion artifacts.
2) With the inclusion of self-assisted priors as well as spatial attention, we develop end-to-end stacked refinement networks. This can further improve the pixel-to-pixel dependency throughout pixel-wise matching and spatial location preserving between predictions.
3) We provide an intensive analysis of using different types of image priors: the proposed self-assisted priors and priors from other image contrast of the same subject. This analysis proves the effectiveness and feasibility of our self-assisted priors since it does not require any further data scans.
4) The source code of our proposed stacked U-Nets with self-assisted priors is available at *https://github.com/Yonsei-MILab/MRI_Motion_Artifact_Correction_Self-Assisted_Priors*.

The rest of this article is organized as follows. Section 2 provides theory and details of the proposed stacked networks. Section 3 presents the motion artifacts simulation and the ablation analysis of using different types of priors. Experimental results of the simulated and in-vivo data are presented in Section 4. Eventually, Section 5 and 6 discuss and conclude the paper, respectively.

## 2. Methodology

### 2.1. Theory

The goal of the correction methods of MRI motion artifacts is to recover the motion-free image $x$ from the motion-corrupted image $x_m$ that was caused due to the patient movement during data acquisition. This motion causes a degradation at the time of acquiring the k-space points. To simplify the problem formulation, we generally defined the image degradation in the spatial domain as follows:

$$x_m = \text{Tr}[x], \qquad (1)$$

where Tr is the rigid motion transformation defined as $\text{Tr} = \mathcal{F}^{-1}\mathcal{M}\mathcal{F}\text{T}_\theta \mathcal{R}_\theta$. $\text{T}_\theta$ and $\mathcal{R}_\theta$ are the translation and rotation operators in the image domain, while $\theta$ indicates their motion parameters $(t_x, t_y, t_z)$ and $(\rho_x, \rho_y, \rho_z)$, respectively. $\mathcal{F}$, $\mathcal{M}$, and $\mathcal{F}^{-1}$ are the Discrete Fourier Transform (DFT), sampling operator in k-space, and inverse DFT, respectively. Basically, it is hard to solve this equation linearly since it could contain many unknowns (i.e., motion parameters). However, this task can be formed as an optimization problem using non-linear deep learning networks to derive the inverse transformation map $\text{Tr}^{-1}$ that leads to obtaining the corrected image $\hat{x}$ as follows:

$$\hat{x} = \underset{x}{argmin} \left\| \text{Tr}^{-1}[x_m] - x \right\|_2^2. \qquad (2)$$

It is assumed that the structural details of the white matter and gray matter in the brain are similar within the contiguous slices, and in particular, during the 3D imaging. Thus, the inclusion of additional prior knowledge from the adjacent slices of the same corrupted subject could assist to significantly solve this problem. The above equation can be re-formulated as:

$$\hat{x}^{[i]} = \underset{x^{[i]}}{argmin} \left\| \text{Tr}^{-1}\left[x_m^{[i-1]} + x_m^{[i]} + x_m^{[i+1]}\right] - x^{[i]} \right\|_2^2, \qquad (3)$$

where $x_m^{[i-1]}$ and $x_m^{[i+1]}$ are the image priors derived from the adjacent slices of the current motion-corrupted image $x_m^{[i]}$. In spite of the fact that the utilized image priors are corrupted data, they can share some details that may be missed in the current $i$th image. It is of note that this is a multi-input single-output process, which corrects the motion artifacts for a single motion image at every time. Therefore, the requirement of learning additional structural information can be addressed through applying the concept of the proposed self-assisted priors.

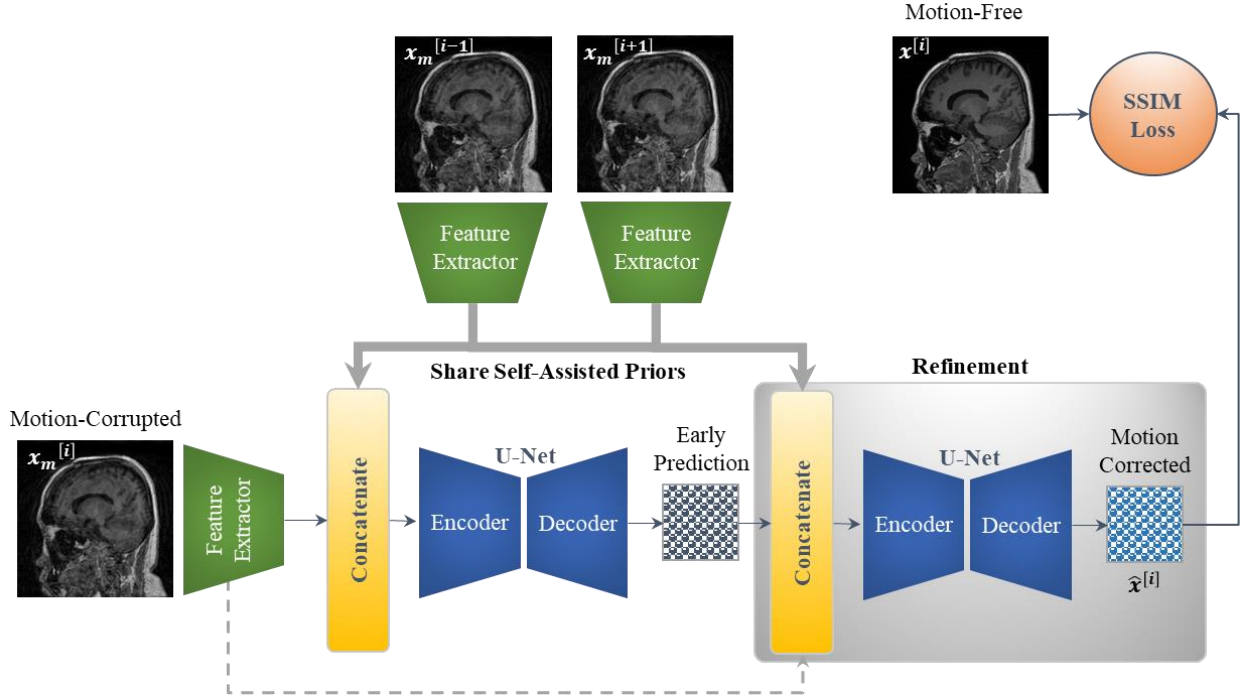

**Figure 1.** Schematic diagram of the proposed stacked U-Nets with self-assisted priors for motion artifact correction.

## 2.2. Proposed Motion Artifact Correction Network

Inspired by U-Net [30], which was at first designed for image segmentation and then was successfully applied to the denoising and reconstruction tasks, we propose stacked U-Nets with self-assisted priors for image correction of the MRI motion artifacts. An overview of the proposed motion correction network is graphically illustrated in Figure 1. The primary goal of designing such a deep learning network is to retrieve the motion-free image and increase the readers' ability to correctly diagnose patients from the structure of interest that does not include any obstacles such as artifacts. The essential contributions of the proposed architecture include both the self-assisted priors and stacked refinement network. The following subsections describe these concepts in detail.

### 2.2.1. Self-Assisted Priors

The proposed network initiates by integrating the image priors from the adjacent slices of the target corrupted image. This is achieved by re-designing the original U-Net to be capable of importing and passing multiple inputs instead of a single input. A grayscale motion-corrupted image $x_m^{[i]}$ along with its self-assisted priors $x_m^{[i-1]}$ and $x_m^{[i+1]}$ with a fixed image size of 256×256 pixels are passed independently to feature extractor encoders, generating 32 separate feature maps for each input. All these extracted features are then concatenated to serve as an input to the proposed network as shown in Figure 1. This process is applicable to all image slices within the training and testing subjects, except for the first and last slices, in which they were repeated due to the absence of the other prior. The inclusion of these image priors assists to share some structural details that may be missed in the target motion-corrupted image $x_m^{[i]}$, resulting in improvements in recovering the motion-free image. This is due to the fact that the brain structure is of high homogeneity in the contiguous locations, especially if the data acquisition were performed in 3D imaging strategy as the case of this work (i.e., no impact of the slice gap). In other words, the 3D imaging produces higher spatial resolution, which makes the usage of additional prior knowledge from the adjacent slices is of great importance to retrieve the missed parts and, hence, achieve better motion correction. It should be noted that this is a multi-input single-output process, which implies that the network learns how to correct only the motion-corrupted image $x_m^{[i]}$ in a supervised manner by computing the loss with the corresponding ground-truth motion-free image $x^{[i]}$. In each training iteration, the role of the priors $x_m^{[i-1]}$ and $x_m^{[i+1]}$ is to feed additional useful patterns from the same corrupted subject into the network. However, correction of priors does not take place during the correction of the current motion-corrupted image $x_m^{[i]}$. Compared to [22, 25], the proposed self-assisted priors approach has the following advantages: (i) it eliminates the need for additional MRI scans to be used as image priors, and (ii) it also reduces the computational cost since it does not require any further image pre-processing such as image registration and alignment. Nevertheless, if another contrast data is available, it could be used as an additional image prior besides the self-assisted priors of the same contrast. In this paper, we test different types of image priors, including both the self-assisted priors as well as the priors from other contrast data. Figure 2 shows some examples of the adjacent slices with their SSIM scores related to the current corrupted slice. The relatively high SSIM scores of the closer adjacent slices ($x_m^{[i-1]}$ and $x_m^{[i+1]}$) in this figure exemplifies the benefits of our assumption of using self-assisted priors for better motion correction by sharing some missed structural details. It is observed that the farther slices ($x_m^{[i-2]}$ and $x_m^{[i+2]}$) have less similarities with the current motion-corrupted image.

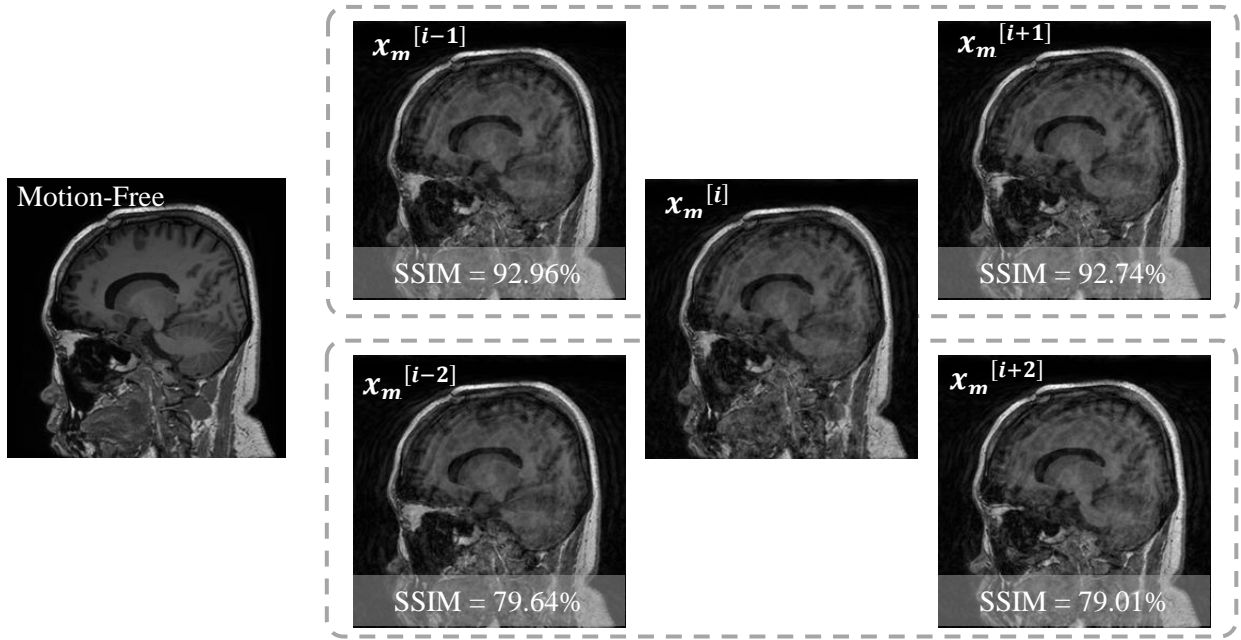

**Figure 2.** Examples of using the adjacent slices ($x_m^{[i-1]}$, $x_m^{[i+1]}$) and ($x_m^{[i-2]}$, $x_m^{[i+2]}$) of a certain corrupted image $x_m^{[i]}$ as additional self-assisted priors with their SSIM scores. The computed SSIM score of $x_m^{[i]}$ with its ground-truth motion-free image is 68.86%.

Therefore, we conducted this work using the self-assisted priors from only the closer adjacent slices since they contain higher similarities.

### 2.2.2. Stacked Refinement U-Nets

The idea of the stacked networks has been utilized in the field of semantic segmentation [31, 32] in order to refine the segmented target and achieve better performance. In this work, we proposed a new architecture consists of two cascaded encoder-decoder U-Nets. We developed the stacked U-Nets with the reuse of image priors throughout all networks. More specifically, the early prediction of the first encoder-decoder U-Net was concatenated along with the image priors and their corresponding motion-corrupted image and passed into the second encoder-decoder network. This connection enables the second network to capture more spatial differences among the two predictions and hence, achieve better refinement of the corrected images. The proposed stacked refinement network was trained in an end-to-end manner. Thus, the initial estimates were re-evaluated during the training process. In the end, the proposed stacked U-Nets with the benefits of sharing additional knowledge could improve the pixel-to-pixel dependency, and leads to recovering motion-free images with significant preservation of spatial information.

### 2.2.3. Attention Module

We also adopted the Convolutional Block Attention Module (CBAM) [33] into all resolution levels of the proposed stacked networks. The CBAM module was achieved by integrating two attention maps along the channel and spatial directions. This can improve the representation of interests and focus more on regions that may need extra correction.

### 2.2.4. Implementation Details

In this work, the proposed network was trained in a supervised manner using the simulated motion-corrupted images with their corresponding ground-truth motion-free data. The exploited U-Net contains four encoder-decoder levels. The convolution kernels throughout all networks have the size of 3×3 with feature maps of 32, 64, 128, and 256 in the encoder path, while reverse feature maps were utilized in the decoder path followed by a final output prediction map. We applied both the batch normalization and activation function of Rectified Linear Unit (ReLU) sequentially over all the convolutional layers. We used the average pooling with a filter size of 2×2 and a stride of 2. It is noteworthy that the feature maps in each resolution level of the encoder path were concatenated with the corresponding maps of the same resolution level at the decoder path. This allows fusing both the global and local representations among early and late convolution layers. During network training, we utilized the Adam optimizer with a batch size of 10 samples for gradient descent. We initially set the learning rate to 0.001 and it is decayed exponentially by a factor of 10 throughout 50 epochs. The proposed network took around 5.8 h to accomplish its training over 4.01 million parameters.

The implementation of this paper was performed on a PC equipped with the following specifications: a Cuda-enabled GPU of NVIDIA GeForce RTX 3080 with 128 GB RAM. This work was implemented using Python 3.7.10 and Tensorflow with Keras library on the Ubuntu 18.04 operating system.

## 2.3. Evaluation Measures

Previous studies for image translation used measures such as the SSIM, Mean Square Error (MSE), Peak Signal-to-Noise Ratio (PSNR), or perceptual losses, however, it is not certain as to which measure would be most sensitive for evaluating motion. While MSE has been utilized as a loss function in many image denoising and reconstruction applications, SSIM loss has been widely preferred in motion correction tasks. This is due to that motion artifact significantly degrades the image structure and the SSIM loss basically computes the relation of the structural similarities among the reference and target images.

Here, we investigated which measurement is more sensitive to different types of motions and then can be utilized as the most effective loss function. To perform this experiment, we generated three different types of motions, namely mild, moderate, and severe for the same patient data [34]. Our analysis concentrated on the SSIM, MSE, PSNR, and Perceptual from the pre-trained Visual Geometry Group Network (VGGNet) indices since they are usually utilized in the denoising tasks. Figure 3 illustrates 3D plots that show the relationship among various motion types for each index. We observed that the SSIM and PerceptualVGG measures provided better indications on distinguishing various motion artifact strengths (i.e., mild, moderate, and severe motions) compared to the MSE and PSNR. It is of note that the PerceptualVGG was computed by passing all the motion-corrupted images and their reference motion-free data into a fixed pre-trained model such as the VGGNet and then quantifying the feature differences among them over a certain layer within the network. In terms of correlations between different motion pairs, the SSIM index shows the best correlations with very high R-squared values of 0.9243 for mild vs. moderate, 0.9667 for mild vs. severe, and 0.9191 for moderate vs. severe. It is also observed that the PerceptualVGG shows a better correlation of different motion types compared to the MSE and PSNR. Hence, this investigation provides a justification for using the SSIM as a loss function.

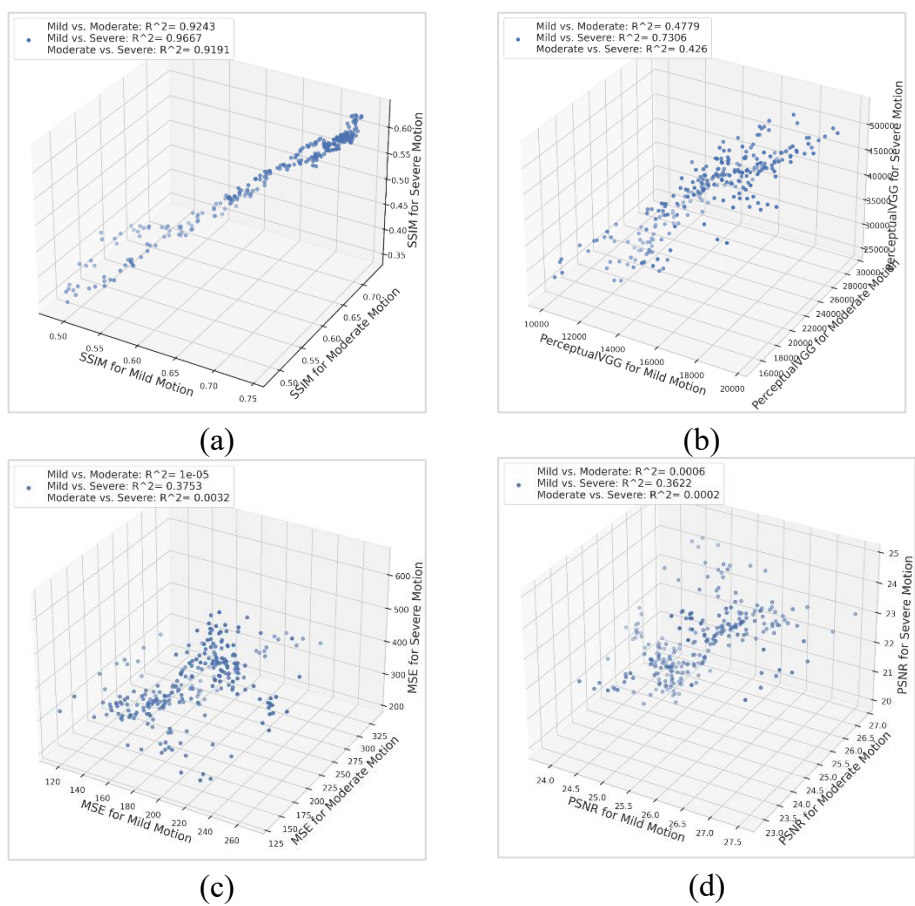

**Figure 3.** Relationship between various motion severity levels in terms of (a) SSIM, (b) PerceptualVGG, (c) MSE, and (d) PSNR. The correlations of various motion pairs are reported using the R-squared values.

## 3. Experiments

### 3.1. Original Dataset

To accomplish the motion artifact correction task, we have collected a set of clinical brain MRI data at Seoul National University Hospital (SNUH). Our dataset contains 83 motion-free clnical subjects, which was utilized to generate the simulated motion data to conduct network training. Additionally, separate 24 clinical subjects with motion distorted were acquired to perform the in-vivo assessment. 3D T1-weighted gradient echoes, called BRAin VOlume (BRAVO), were acquired on a 3.0 Tesla MRI scanner (SIGNA Premier, GE Healthcare, United States) with the following imaging parameters: echo time (TE) of 2.77 ms, repetition time (TR) of 6.86 ms, flip angle of 10º, and pixel bandwidth of 244 Hz/pixel. The sagittal image matrix varies from 256×256 to 512×512 pixels with an in-plane resolution of 0.90×0.90 $mm^2$ and slice spacing of 1 mm, while the third dimension is in the range of 144 to 380.

### 3.2. Simulation of Motion Artifacts

Since there is a limitation to acquire various pairs of reference motion-free and target motion-corrupted datasets, simulation of MRI motion artifacts is inevitable to fulfill the network training. This was achieved by synthesizing the motion-corrupted images from the 83 original motion-free subjects. In this work, 3D rigid brain motion was simulated to generate the motion-corrupted data. To achieve the feasibility and reliability of the proposed motion correction method, it is important to generate motion data that highly imitates the real motion artifacts. To do so, we assumed that the bulk rigid motion is a combination of both the rotation and translation motions. Motion artifacts were generated by applying sporadic rotational motions in range of [–7º, +7º] on all three axes as well as by applying translational motions between −7 and +7 mm on all three planes. As aforementioned in the theory section, both the rotation and translation motions could be achieved by controlling their parameters $(\rho_x, \rho_y, \rho_z)$ and $(t_x, t_y, t_z)$. To generate various motion severity (i.e., various strengths of motion artifacts), similar to the real motion cases, the motion parameters were acquired by generating random peaks of various numbers and ranges. It is noteworthy that this motion simulation could be performed in the image domain or frequency domain. In this work, we performed a complicated motion simulation process simultaneously in both image and frequency domains in order to derive the simulated motion data with very high similarity to the real motion cases. More specifically, the rotation motion was performed in the image domain (image-based simulation) by applying some rotational operations on a motion-free image and then sampling the k-space lines of the Fourier transform of that rotated image. Meanwhile, the translation motion (k-space-based simulation) was performed by adding linear phase shifts directly to the k-space in the phase-encoding direction. Figure 4 demonstrates the entire motion simulation process. More details about the motion simulation tool is available in our previously published paper [35].

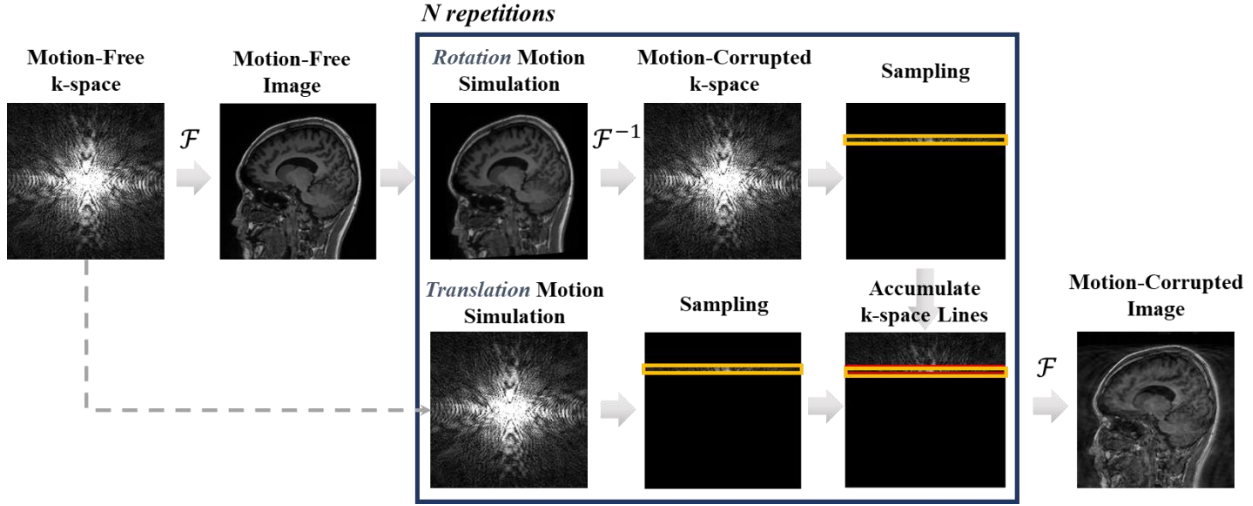

**Figure 4.** Whole process of the motion artifacts simulation. Both the rotational and translational operators were utilized and proceeded in the image domain and k-space domain, respectively.

To accomplish the network training and evaluation, we divided the synthesized motion data based on the subject level into two main sets, namely the training and testing with portions of 80% and 20%, respectively. In term of images, a total of 9,996 images were utilized for training, while 3,390 images were used for testing. Among the training data, we randomly selected 5% of the images and used them for validation.

### 3.3. Ablation Studies

The goal of this section is to experimentally investigate different ablation studies that make the proposed network powerful enough to correct the motion artifacts from MR images, such as the inclusion of image priors from the same corrupted subjects and the incorporation of additional data from different image contrast. Throughout all these investigations, we have monitored the motion correction performance using the SSIM and MSE indices.

### 3.3.1. Image Priors from Same Corrupted Subjects

This section presents the ablation experiments for the main contributions of this paper. As illustrated in Table 1, the motion simulation tool produced motion-corrupted images with a degradation in the image quality of 71.66%, 99.25, and 28.83 in terms of SSIM, MSE, and PSNR, respectively. The first experiment was straightforward by passing a single corrupted image into the original supervised U-Net. The network learned how to reduce the motion artifacts by measuring the SSIM loss between the input corrupted images $x_m^{[i]}$ and its reference motion-free images $x^{[i]}$. The results of this experiment showed an improvement in the motion correction compared to the corrupted simulated data with SSIM of 94.20% and MSE of 37.06%. In the second experiment, we included additional knowledge from the adjacent slices ($x_m^{[i-1]}$ and $x_m^{[i+1]}$) of the same corrupted image into the network. We called this process self-assisted priors since all the incorporated information was derived from the same patient. This yields a network with a multi-input single output structure, which implies that at each iteration the network takes the corrupted image and its adjacent slices as inputs and be able to construct and correct the motion of only the current slice. This experiment could achieve a slight improvement rate of 0.22% in term of SSIM, while it obtained a significant reduction of MSE from 37.06 to 33.87. Next, we examined the effect of adding the CBAM attention module into the network. We could gain some improvements in the overall performance of the motion correction with an SSIM score of 94.64%. In the last experiment, we trained and tested all the above components via the stacked U-Net (see Figure 1). The main purpose of this stacked architecture was to get better refinement of the motion correction image. The self-assisted priors were also shared with the second stacked network. As presented in Table 1, the proposed stacked U-Nets with the inclusion of the self-assisted priors achieved significant improvements with SSIM of 95.03%, MSE of 29.76, and PSNR of 33.81.

**Table 1.** Experimental results of the proposed stacked U-Nets with the self-assisted priors compared to the corrupted simulation results

| Experiment | SSIM (%) | MSE | PSNR |
| --- | --- | --- | --- |
| *Corrupted Simulated Data* | *71.66* | *99.25* | *28.83* |
| Single U-Net with Single Corrupted Input (No Priors) | 94.20 | 37.06 | 32.87 |
| Single U-Net with Inclusion of Self-Assisted Priors | 94.44 | 33.87 | 33.27 |
| Single U-Net with Inclusion of Self-Assisted Priors and Attention Module | 94.64 | 33.85 | 33.25 |
| Stacked U-Nets with Inclusion of Self-Assisted Priors and Attention Module | **95.03** | **29.76** | **33.81** |

### 3.3.2. Additional Prior from Different Contrast-Enhanced (CE) Data

Despite the cost, time, and patient inconvenience, the acquisition of additional MRI scans could be crucial in some clinical applications, especially the CE images. Fortunately, there were 38 patients among our dataset that have additional CE data with the same imaging parameters. This section exhibits the importance of using the CE data as an additional image prior besides the self-assisted priors. To accomplish this experiment, we have split this new dataset based on the subject level into 80% for training (3,684 images) and 20% for testing (1,378 images). Here, we only used the 38 simulated motion-corrupted data from the T1-weighted BRAVO subjects with their corresponding CE data. Thus, this is a smaller dataset compared to the previous experiments. Table 2 presents the effect of using additional CE data as an image prior besides the self-assisted priors. The image quality of this simulated motion-corrupted dataset was reduced to 68.54%, 123.23, and 27.48 in terms of SSIM, MSE, and PSNR, respectively. All the investigations in this section were performed utilizing the stacked U-Net architecture. In order to examine the effect of using different types of image priors, we started training the stacked networks without incorporating any image priors (i.e., neither the self-assisted priors nor the CE data prior). This stacked network without any image priors reduced the motion artifacts and obtained the SSIM score of 91.77% and MSE of 62.96. However, significant improvement was achieved by the inclusion of only the self-assisted priors. The performance was improved in term of SSIM from 91.77% to 92.10%, while the MSE was dramatically reduced in term of MSE from 62.96 to 53.55. It is noteworthy that for the CE prior, we utilized the same image slice $CE^{[i]}$ of the target motion-corrupted image $x_m^{[i]}$. Also, promising achievement was obtained in the case of using only the CE data as image prior. It improved the performance with overall increments in term of the SSIM from 91.77% to 93.03% and decrements in term of MSE from 62.96 to 59.62. It is observed that the inclusion of self-assisted priors achieved a better MSE of 53.55 compared to the using of CE prior, which obtained 59.62. In opposite, the performance of the proposed network with the CE prior outperformed the network with the inclusion of the self-assisted priors with scores of 93.03% and 92.10% in terms of SSIM, respectively. In the last experiment, we integrated both priors together. Thus, to correct the target motion-corrupted image $x_m^{[i]}$, we used four inputs into the network. Three of them are the image priors ($x_m^{[i-1]}$, $x_m^{[i+1]}$, and $CE^{[i]}$) and last one is the target image itself ($x_m^{[i]}$). This approach results in maintaining the highest SSIM score of 93.04% and keeping a MSE of 54.06. We conclude that if additional MRI scans are available, they can be used as image priors besides the self-assisted priors to further enhance the performance of motion artifacts correction.

**Table 2.** The effect of using additional CE data as an image prior besides the self-assisted priors

| Experiment | SSIM (%) | MSE | PSNR |
|---|---|---|---|
| *Corrupted Simulated Data* | *68.54* | *123.23* | *27.48* |
| Stacked U-Nets without any Image Prior | 91.77 | 62.96 | 30.43 |
| Stacked U-Nets with only Self-Assisted Priors | 92.10 | **53.55** | **31.06** |
| Stacked U-Nets with only CE Data as Prior | 93.03 | 59.62 | 30.68 |
| Stacked U-Nets with Integrated Self-Assisted Priors and CE Data Prior | **93.04** | 54.06 | 31.04 |

## 4. Results

### 4.1. Motion Correction Results on Simulated Dataset

This section shows the motion artifacts correction performance of the proposed stacked U-Nets using two testing groups. The first testing set includes only the image prior from the same corrupted subjects (i.e., the proposed self-assisted priors from the adjacent slices). We evaluated this category using 3,390 corrupted simulated images. The second testing group involves the self-assisted priors with the incorporation of additional prior from different CE data, which has a total of 1,378 corrupted images. As presented in Table 1 and Table 2, the proposed deep learning network was able to learn additional knowledge of the motion patterns from various image priors, leading to achieve promising results in the motion artifacts correction task. For the first testing group, the proposed network provides promising results in motion artifacts correction with significant improvement rates of 23.37%, 70.02%, and 17.27% in terms of SSIM, MSE, and PSNR compared to the simulated motion-corrupted data, respectively. Furthermore, the proposed stacked network with the integration of both the self-assisted priors and CE prior achieves promising results in motion artifacts correction with significant improvement rates of 24.50%, 56.13%, and 12.95% in terms of SSIM, MSE, and PSNR compared to the simulated motion-corrupted data, respectively.

Figure 5 illustrates some exemplar motion correction results of the proposed network compared to the ground-truth motion-free images from both testing groups. This figure clearly shows how the proposed method can significantly improve the image quality and reduce the motion artifacts. The zoomed regions of the grey rectangle that was inscribed on the reference image intelligibly revealed the capability of the proposed network to retrieve the corrupted structural details and provide comparable visual performances with the reference motion-free image. Thus, our findings seem to be feasible in increasing the reader's ability to examine and identify the presence of pathologies. One interesting observation from the difference maps in Figure 5 is that the background regions of the corrected images were near totally recovered, while the motions over the brain parenchyma regions were notably corrected better than the skeletal structure of the head. Moreover, we illustrate the motion correction performances for each testing image throughout both testing groups in Figure 6. These boxplots show the SSIM, MSE, and PSNR indices before and after correction of motion artifacts.

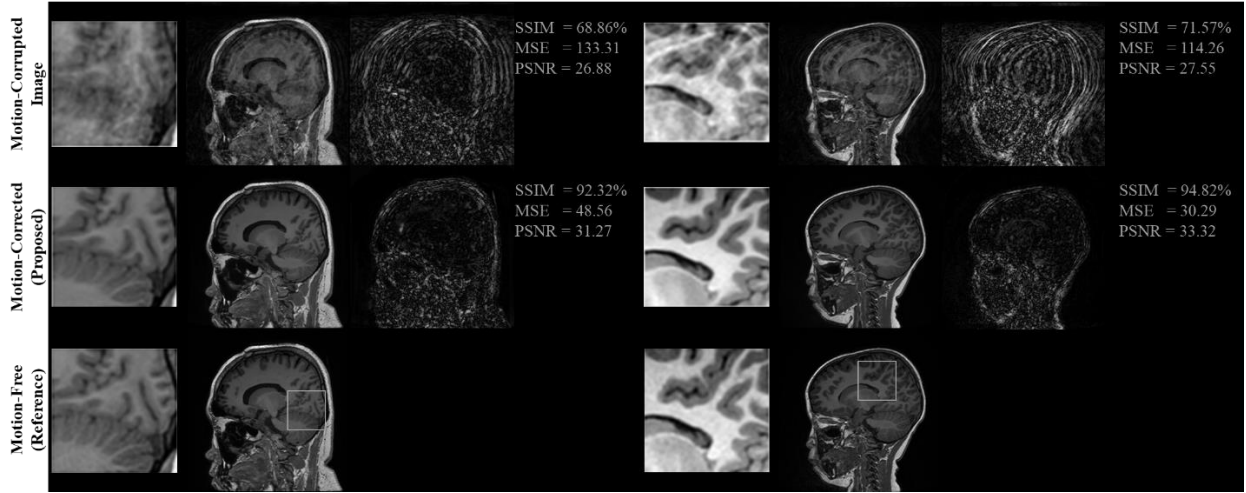

(a)

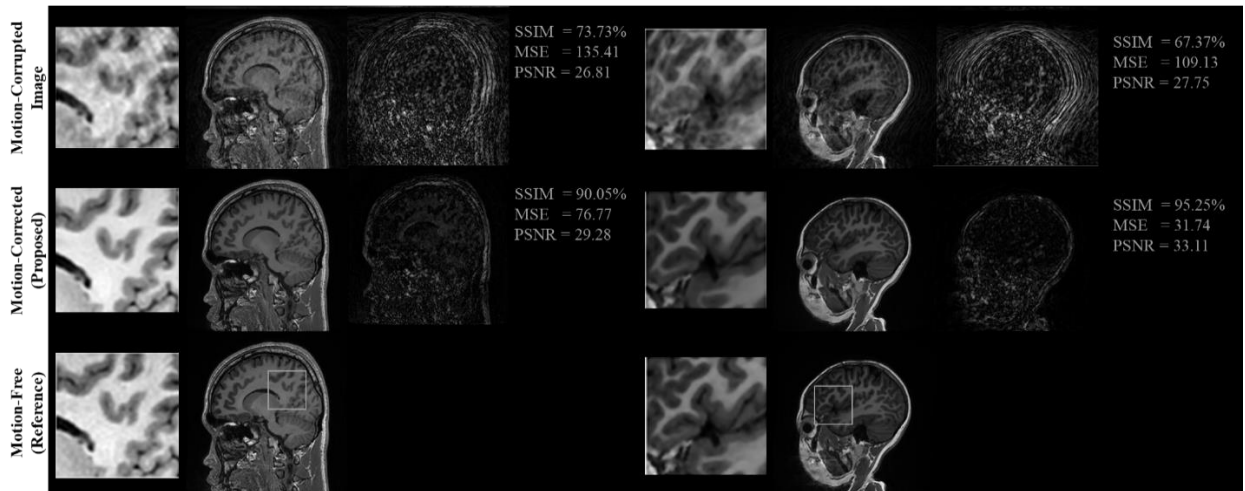

(b)

**Figure 5.** Motion artifacts correction results of the proposed stacked U-Nets with (a) self-assisted priors and (b) integration of self-assisted priors with CE prior. First column of each example shows the zoomed region of the grey rectangle that was inscribed on the reference image. Second column presents the motion-corrupted image, motion-corrected via proposed work, and the reference motion-free image. In the last column of each sample, we show the pixel-wise difference maps compared to the reference image. For each case, we report the evaluation indices before and after the correction of motion artifacts.

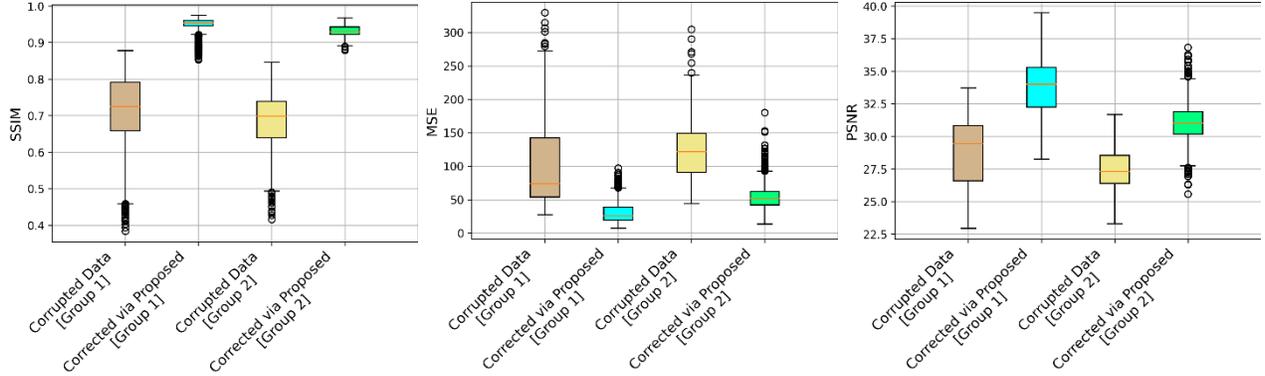

**Figure 6.** Boxplots of the motion correction performances for each testing group. Group 1 refers to the results of the proposed network using the self-assisted priors, while Group 2 indicates the results using additional prior from different CE data. Plots from left to right represent the SSIM, MSE, and PSNR scores, respectively.

### 4.2. In Vivo Experiments: Motion Correction Results on Real Patient Data

To examine the efficiency and feasibility of the proposed motion artifacts correction method, we have tested the network on real clinical motion data. We show some real motion cases in Figure 7. For these cases, clinicians recommended performing another data acquisition due to the poor image quality derived in the first scans that were affected by motion artifacts. As illustrated in this figure, the second scans were more clear data, but with a slight misalignment with the motion distorted images. However, if we assumed that the second scans are the references (or semi-reference) of the distorted images, then the proposed network with self-assisted priors was able to reduce the motion artifacts and obtain improvement rates of 3.68%, 4.90%, 6.83%, and 7.25% in term of SSIM for all presented samples in this figure, respectively. It is noteworthy that the last example in this figure is CE data, while our network was only trained using non-CE data. Even though, the proposed method still achieved promising results in motion artifact correction.

Further, Figure 8 presents the results of motion correction for some examples with various strengths of motion artifacts. For this kind of assessment, it is impossible to compute the quantitative indices due to the absence of the reference motion-free images. Nevertheless, the motion artifacts seem significantly reduced and can be observed visually in the motion-corrected images. The first two examples in Figure 8 show how efficient the proposed network is in retrieving clear brain structures from the real blurred images. In the third example, despite that the real distorted subject has a severe motion artifact, the proposed network could improve the image quality and reduce the motion artifacts. In the last example, it is observed that the acquired real data has extremely severe motion, which results in degrading the image quality to the level of destroying all the brain structural details. In spite of that, the motion-corrected image seems improved in some regions and some structural patterns have appeared.

## 5. Discussion

In this work, we developed a new supervised deep learning network called stacked U-Nets with self-assisted priors towards rigid motion artifacts correction. The proposed network secured two main contributions. The adjacent slices of the corrupted image enable the employment of prior knowledge. These self-assisted priors contain useful information that leads to enhancing the image quality. Additionally, the stacked U-Nets provide a refinement stage, which offers prediction re-estimation and eventually preserves the spatial details. To accomplish the network training, it was inevitable to simulate MRI motion artifacts from motion-free data. This is due to the difficulty of acquiring different pairs of both motion-free and motion-corrupted data from the same patients. As presented in Figure 4, we synthesized the motion-corrupted data using both rotational and translational motions simultaneously in the image and k-space domains, respectively. This procedure can produce various levels of motion severity similar to the real motion data by controlling the motion parameters.

The experiments in this work were conducted to investigate the robustness of the proposed network to enhance the motion artifacts correction from brain MRI data. The experimental results shown in Table 1 and Table 2 demonstrated the significance of incorporating the additional priors information to improve the overall performance of motion correction. These inclusions of image priors from the adjacent slices of the same corrupted subjects or from other image contrasts (i.e., CE data in our work) enable the network to share some missing structural patterns such as borders of the white and grey matters in the brain. This idea offers a potential solution to correct the MRI motion artifacts with a significant improvement compared to the simulated corrupted data. For the case of using only the self-assisted priors from the corrupted data itself, the overall performance was significantly enhanced. We also achieved better improvements in the correction of motion artifacts in the case of incorporating both self-assisted priors with the additional CE prior data. In this case, the overall SSIM and MSE scores were improved. These results indicate the high similarity of the anatomical structure of the brain in the corrected image compared to its motion-clean image. In both cases, the proposed motion correction method quantitatively outperformed the deep learning-based U-Net.

We also inspected the data consistency of the motion-corrected in other axial and coronal planes, as illustrated in Figure 9 from an actual clinical case. It is clear that the reconstructed motion-corrected volumes maintain the uniformity of the brain structure while reducing the motion artifacts. Note that viewing from different planes show interslice dependencies in the distorted images. The corrected images do alleviate this, but there is still some interslice differences. This could be a limitation of the study. Generally, the proposed self-assisted priors have the ability to share useful patterns from the contiguous slices and assist to retrieve better image quality. Other image contrasts such as CE data are of high accessibility in realistic clinical exams. Thus, employing both image priors (i.e., self-assisted priors and extra image contrast priors) of the same subject could improve the performance of artifact motion correction.

A comparison against the latest studies in the literature has been reported in Table 3. As presented, each work had generated simulated input data with various levels of motion severity and utilized different amounts of training and testing MRI subjects. Nevertheless, all methods achieved improvements in the reduction of motion artifacts from MRI images. It is observed that the work in [25] generated simulated motion data with mild motion, and hence obtained lower improvement rates throughout three different image contrasts compared to other works listed in this table. We also compared the proposed method against the original U-Net [30] and Cycle Generative Adversarial Networks (CycleGAN) [36] using the same training and testing sets. The results showed that the proposed network outperformed the U-Net and CycleGAN with overall improvement rates of 23.37% vs. 22.54% and 20.13%, 70.02% vs. 62.66% and 46.12%, and 17.27% vs. 14.01% and 7.11% in terms of SSIM, MSE, and PSNR, respectively. Although the CycleGAN obtained remarkable success in the image-to-image translation tasks, it showed limited achievements to address the problem of motion artifact correction, especially if the level of motion in the corrupted image is high. A qualitative comparison is illustrated in Figure 10. It is shown how efficient the proposed network is in producing high image quality that highly matches with reference motion-free data. In contrast, CycleGAN struggled to reduce motion artifacts in the severe motions, as presented in the first two examples of Figure 10. In these instances, CycleGAN generated results with some structural deterioration. However, CycleGAN obtained better motion correction performances in the cases of moderate motions compared to severe ones, as presented in the last two examples. Nevertheless, the proposed method still overcomes the CycleGAN and U-Net on different motion levels. An interesting observation is that the motion correction via U-Net achieved better performance compared to the CycleGAN method.

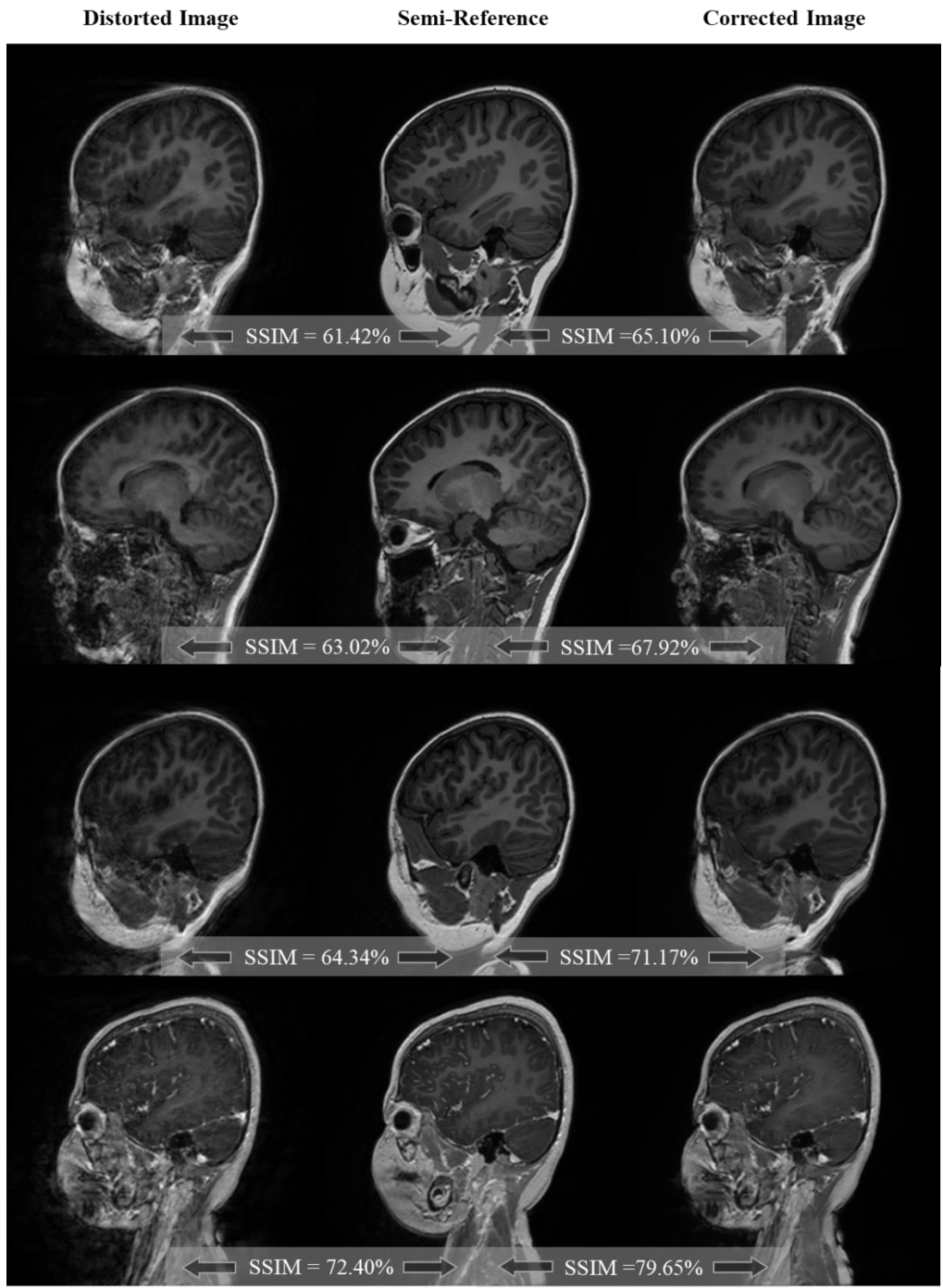

**Figure 7.** Real motion cases. From left to right, the distorted real clinical motion images, the semi-reference image of the same distorted data (another scan of the same patient), and the motion-corrected images via the proposed stacked network with the self-assisted priors.

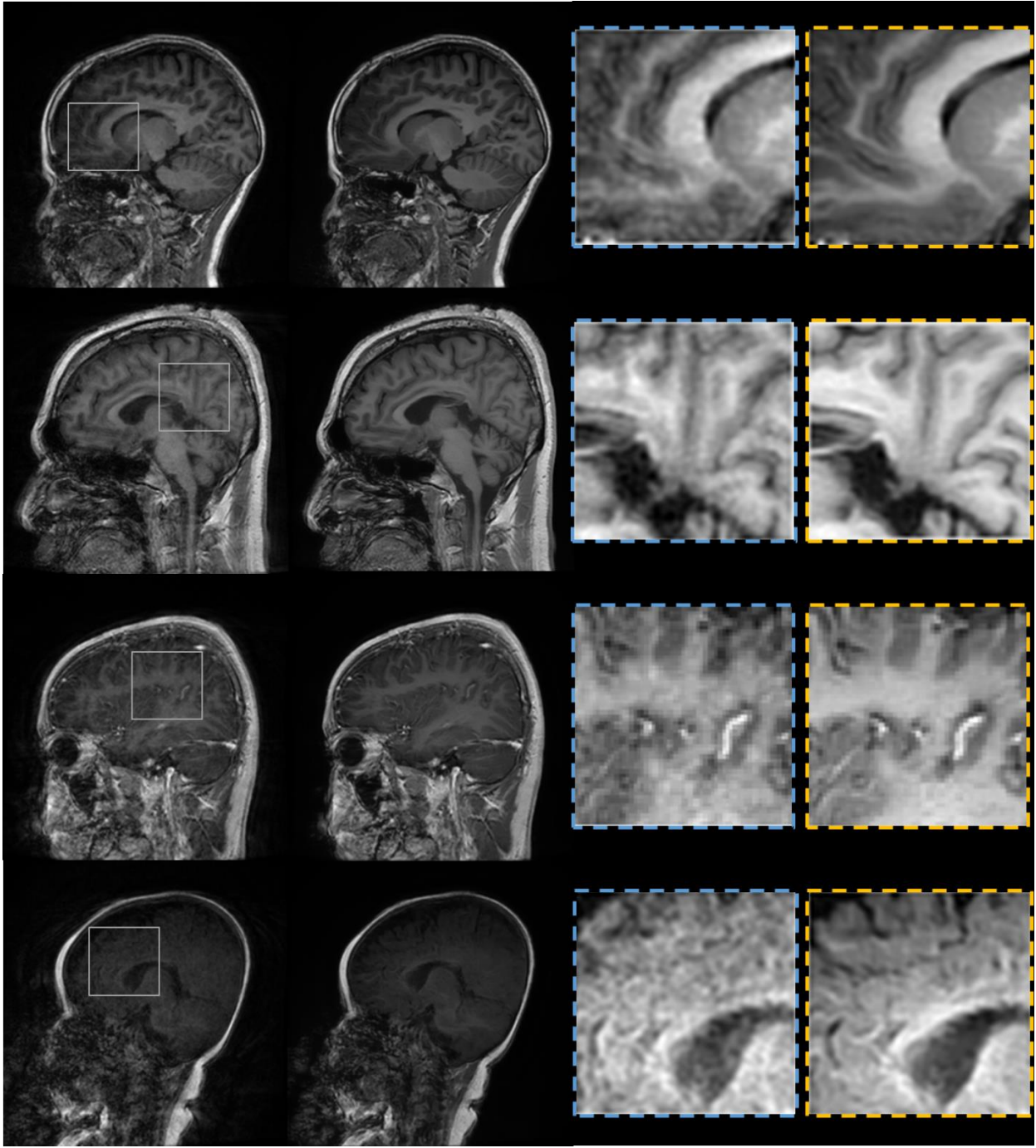

**Figure 8.** Results of motion artifacts correction for four clinical subjects. First column indicates the real images with motion artifacts, while second column represents the motion-corrected images via the proposed stacked U-Nets with self-assisted priors. The blue and yellow boxes in the right highlight the zoomed regions of both real motion-distorted and motion-corrected via proposed network, respectively.

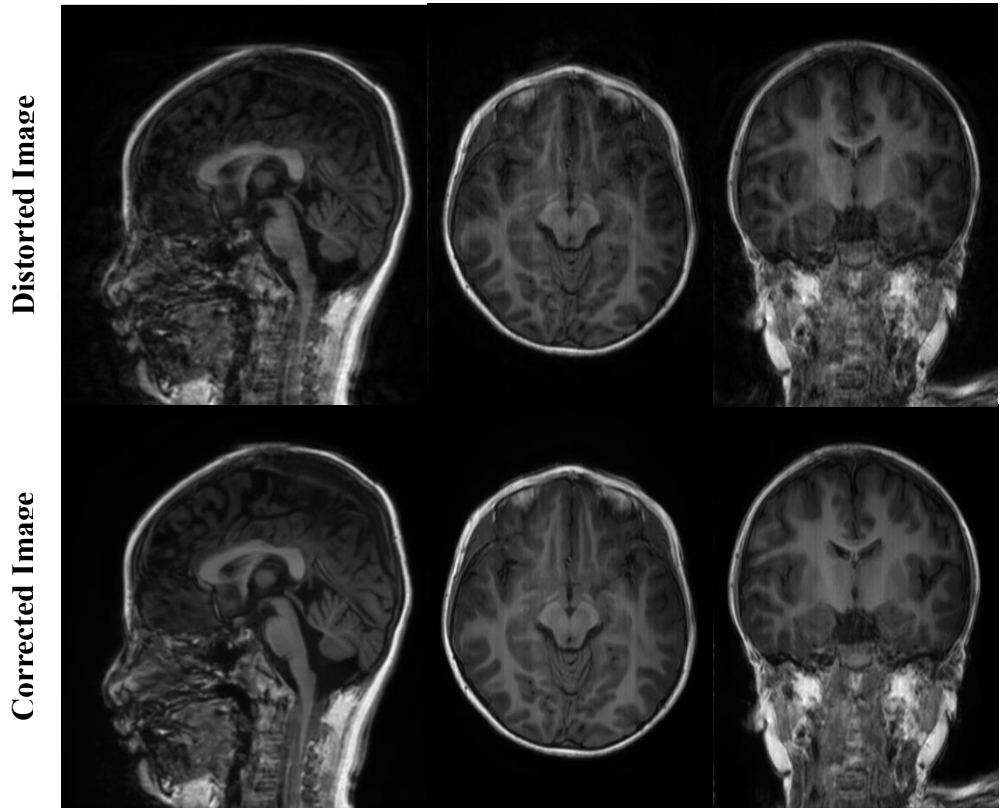

**Figure 9.** Performance of motion correction results on the other axial and coronal planes. It is of note that the network was trained and tested using only the sagittal images. This figure shows the central planes of a real clinical patient.

The limitations of this work are as follows. First, even though the proposed self-assisted priors' strategy was beneficial to improve the motion correction by sharing some missing structural details, the utilized adjacent slices were derived from the same corrupted data. That implies there is still a loss of complete information. Due to this, most motion correction works cannot resolve this problem completely. Thus, we attempted to learn our proposed network using more additional knowledge throughout including the CE image prior. It may be interesting if multiple images of the same subject can be used as additional image priors besides the self-assisted priors for further improvement of motion artifacts correction. However, there is a tradeoff since this requires additional cost by acquiring new MRI scans. Second, it seems there is a difficulty to recover the very severe motion cases with severe blurring or ghosting. An exemplar with a very high motion severity is illustrated in Figure 8 (last example). A complete loss of the borders of the white and grey matters can be clearly noticed, making it a challenging case for any motion correction method. Such cases may lead to an inaccurate diagnosis, and therefore the MRI scan must be repeated. Third, there is an absence of paired real motion-free and repeated motion-corrupted scans. All the motion artifacts correction works have trained their networks using simulated motion data. An interesting work could be to collect an open-access database to address this research topic.

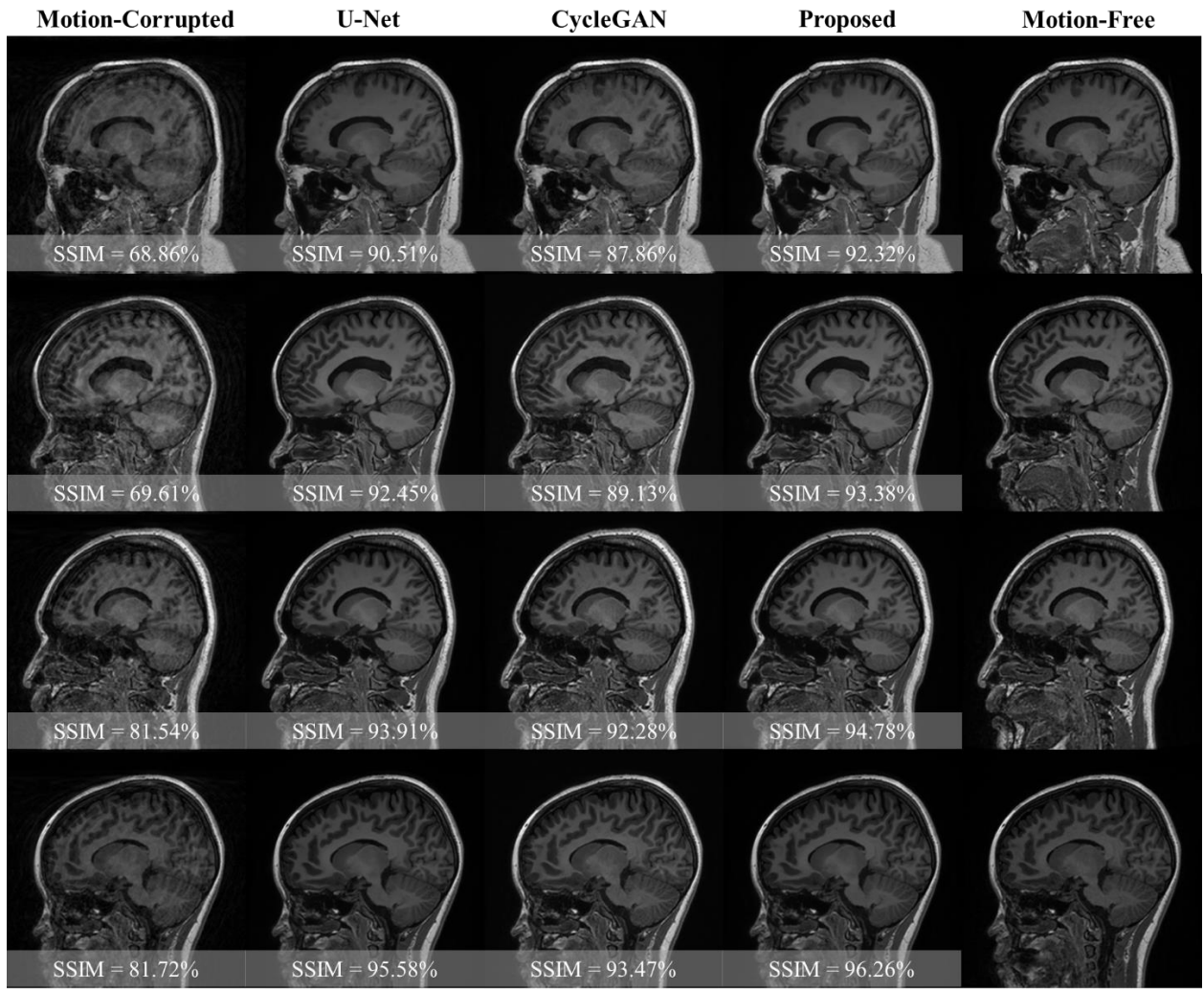

**Figure 10.** Qualitative comparison of the proposed stacked U-Nets with self-assisted priors against the U-Net and CycleGAN methods. All the SSIM values inscribed on the images are computed corresponding to the reference motion-free images.

**Table 3.** Comparison of the proposed stacked U-Nets with self-assisted priors against the latest works in the literature on MRI motion artifact correction. The measures in parenthesis indicate the improvement rates

| Reference | Subjects | Corrupted | | | Corrected | | |
|---|---|---|---|---|---|---|---|
| | | SSIM | MSE | PSNR | SSIM | MSE | PSNR |
| Wang et al. [28] 2D Motion Correction | 1113 (Sagittal) | 40.0. | - | 18.50 | 89.00 (49.00%) | - | 29.60 (60.00%) |
| Wang et al. [28] 3D Motion Correction | | 74.00 | - | 27.20 | 89.00 (15.00%) | - | 32.40 (19.12%) |
| Liu et al. [5] | 85 (Multi-Planes) | 86.7 | - | - | 96.5 (9.80%) | - | - |
| Chatterjee et al. [22] | 100 (Axial) | 77.00 | - | - | 97.00 (20%) | - | - |
| Lee et al. [25] ($T_{1w}$) | | 98.11 | - | - | 99.05 (0.94%) | - | - |
| Lee et al. [25] ($T_{2w}$) | 41 (Axial) | 91.25 | - | - | 98.21 (6.96%) | - | - |
| Lee et al. [25] (FLAIR) | | 92.79 | - | - | 97.48 (4.69%) | - | - |
| U-Net [30] (Implemented) | 83 (Sagittal) | 71.66 | 99.25 | 28.83 | 94.20 (22.54%) | 37.06 (62.66%) | 32.87 (14.01%) |
| CycleGAN [36] (Implemented) | | | | | 91.79 (20.13%) | 53.48 (46.12%) | 30.88 (7.11%) |
| **Proposed (Self-assisted Priors)** | | | | | 95.03 (23.37%) | 29.76 (70.02%) | 33.81 (17.27%) |
| **Proposed (Additional CE Prior)** | 38 (Sagittal) | 68.54 | 123.23 | 27.48 | 93.04 (24.50%) | 54.06 (56.13%) | 31.04 (12.95%) |

## 6. Conclusion

This paper presents a deep learning motion artifact correction method called stacked U-Nets with self-assisted priors. The proposed network exploits the employing of the adjacent slices of each corrupted image to learn additional knowledge that may contain some useful structural details. Specifically, the 3D imaging enables to construct higher spatial resolution, which makes the usage of additional prior knowledge from the adjacent slices is of significance in retrieving the missed parts. Further, we design a refinement stage via developing the stacked U-Nets, which facilitates the generation of better motion-corrected images with superior maintaining of the image details and contrast. Generally, the proposed method increases the efficiency of the clinicians to correctly diagnose MR images by improving the overall clinical image quality. We conclude that if additional MRI scans are available, they can be used as image priors besides the self-assisted priors to further enhance the performance of motion artifacts correction.

## Acknowledgments

This work was supported in part by GE Healthcare research funds.

## References

[1] M. Zaitsev, J. Maclaren, and M. Herbst, "Motion artifacts in MRI: A complex problem with many partial solutions," *Magn Reson Imaging,* vol. 42, no. 4, pp. 887-901, Oct, 2015.

[2] J. B. Andre, B. W. Bresnahan, M. Mossa-Basha, M. N. Hoff, C. P. Smith, Y. Anzai, and W. A. Cohen, "Toward Quantifying the Prevalence, Severity, and Cost Associated With Patient Motion During Clinical MR Examinations," *J Am Coll Radiol,* vol. 12, no. 7, pp. 689-95, Jul, 2015.

[3] F. Godenschweger, U. Kagebein, D. Stucht, U. Yarach, A. Sciarra, R. Yakupov, F. Lusebrink, P. Schulze, and O. Speck, "Motion correction in MRI of the brain," *Phys Med Biol,* vol. 61, no. 5, pp. R32-56, Mar 7, 2016.

[4] R. Shaw, C. H. Sudre, T. Varsavsky, S. Ourselin, and M. J. Cardoso, "A k-Space Model of Movement Artefacts: Application to Segmentation Augmentation and Artefact Removal," *IEEE Trans Med Imaging,* vol. 39, no. 9, pp. 2881-2892, Sep, 2020.

[5] J. Liu, M. Kocak, M. Supanich, and J. Deng, "Motion artifacts reduction in brain MRI by means of a deep residual network with densely connected multi-resolution blocks (DRN-DCMB)," *Magn Reson Imaging,* vol. 71, pp. 69-79, Sep, 2020.

[6] M. D. Tisdall, A. T. Hess, M. Reuter, E. M. Meintjes, B. Fischl, and A. J. van der Kouwe, "Volumetric navigators for prospective motion correction and selective reacquisition in neuroanatomical MRI," *Magn Reson Med,* vol. 68, no. 2, pp. 389-99, Aug, 2012.

[7] M. Herbst, J. Maclaren, M. Weigel, J. Korvink, J. Hennig, and M. Zaitsev, "Prospective motion correction with continuous gradient updates in diffusion weighted imaging," *Magn Reson Med,* vol. 67, no. 2, pp. 326-38, Feb, 2012.

[8] D. Atkinson, H. Derek LG, N. S. Peter, E. S. Paul, and F. K. Steven, "Automatic correction of motion artifacts in magnetic resonance images using an entropy focus criterion," *IEEE Trans Med Imaging,* vol. 16, no. 6, pp. 903-910, 1997.

[9] G. Vaillant, C. Prieto, C. Kolbitsch, G. Penney, and T. Schaeffter, "Retrospective Rigid Motion Correction in k-Space for Segmented Radial MRI," *IEEE Trans Med Imaging,* vol. 33, no. 1, pp. 1-10, Jan, 2014.

[10] A. Loktyushin, H. Nickisch, R. Pohmann, and B. Scholkopf, "Blind multirigid retrospective motion correction of MR images," *Magn Reson Med,* vol. 73, no. 4, pp. 1457-68, Apr, 2015.

[11] M. W. Haskell, S. F. Cauley, and L. L. Wald, "TArgeted Motion Estimation and Reduction (TAMER): data consistency based motion mitigation for MRI using a reduced model joint optimization," *IEEE Trans Med Imaging,* vol. 37, no. 5, pp. 1253-1265, 2018.

[12] M. Talo, O. Yildirim, U. B. Baloglu, G. Aydin, and U. R. Acharya, "Convolutional neural networks for multi-class brain disease detection using MRI images," *Comput Med Imaging Graph,* vol. 78, pp. 101673, Dec, 2019.

[13] D. R. Nayak, R. Dash, and B. Majhi, "Automated diagnosis of multi-class brain abnormalities using MRI images: a deep convolutional neural network based method," *Pattern Recognition Letters,* vol. 138, pp. 385-391, 2020.

[14] M. A. Al-Masni, W. R. Kim, E. Y. Kim, Y. Noh, and D. H. Kim, "Automated detection of cerebral microbleeds in MR images: A two-stage deep learning approach," *Neuroimage Clin,* vol. 28, pp. 102464, 2020.

[15] H. Chung, K. M. Kang, M. A. Al-Masni, C. Sohn, Y. Nam, K. Ryu, and D. Kim, "Stenosis Detection From Time-of-Flight Magnetic Resonance Angiography via Deep Learning 3D Squeeze and Excitation Residual Networks," *IEEE Access,* vol. 8, pp. 43325-43335, 2020.

[16] M. A. Al-Masni, W. R. Kim, E. Y. Kim, Y. Noh, and D. H. Kim, "3D Multi-Scale Residual Network Toward Lacunar Infarcts Identification From MR Images With Minimal User Intervention," *IEEE Access,* vol. 9, pp. 11787-11797, 2021.

[17] B. H. Menze, A. Jakab, S. Bauer, J. Kalpathy-Cramer, K. Farahani, J. Kirby, Y. Burren, N. Porz, J. Slotboom, R. Wiest, and L. Lanczi, "The Multimodal Brain Tumor Image Segmentation Benchmark (BRATS)," *IEEE Trans Med Imaging,* vol. 34, no. 10, pp. 1993-2024, 2015.

[18] M. A. Al-Masni, and D. H. Kim, "CMM-Net: Contextual multi-scale multi-level network for efficient biomedical image segmentation," *Sci Rep,* vol. 11, no. 1, pp. 10191, May 13, 2021.

[19] C. M. Hyun, H. P. Kim, S. M. Lee, S. Lee, and J. K. Seo, "Deep learning for undersampled MRI reconstruction," *Phys Med Biol,* vol. 63, no. 13, pp. 135007, Jun 25, 2018.

[20] K. Ryu, N. Y. Shin, D. H. Kim, and Y. Nam, "Synthesizing T1 weighted MPRAGE image from multi echo GRE images via deep neural network," *Magn Reson Imaging,* vol. 64, pp. 13-20, Dec, 2019.

[21] M. L. Terpstra, M. Maspero, F. d'Agata, B. Stemkens, M. P. W. Intven, J. J. W. Lagendijk, C. A. T. van den Berg, and R. H. N. Tijssen, "Deep learning-based image reconstruction and motion estimation from undersampled radial k-space for real-time MRI-guided radiotherapy," *Phys Med Biol,* vol. 65, no. 15, pp. 155015, Aug 7, 2020.

[22] S. Chatterjee, A. Sciarra, M. Dünnwald, S. Oeltze-Jafra, A. Nürnberger, and O. Speck, "Retrospective Motion Correction of MR Images using Prior-Assisted Deep Learning," in 34th Conference on Neural Information Processing Systems (NeurIPS 2020), Vancouver, Canada, 2020.

[23] M. W. Haskell, S. F. Cauley, B. Bilgic, J. Hossbach, D. N. Splitthoff, J. Pfeuffer, K. Setsompop, and L. L. Wald, "Network Accelerated Motion Estimation and Reduction (NAMER): Convolutional neural network guided retrospective motion correction using a separable motion model," *Magn Reson Med,* vol. 82, no. 4, pp. 1452-1461, Oct, 2019.

[24] Y. Ko, S. Moon, J. Baek, and H. Shim, "Rigid and non-rigid motion artifact reduction in X-ray CT using attention module," *Medical Image Analysis,* vol. 67, pp. 101883, 2021/01/01/, 2021.

[25] J. Lee, B. Kim, and H. Park, "MC(2)-Net: motion correction network for multi-contrast brain MRI," *Magn Reson Med,* vol. 86, no. 2, pp. 1077-1092, Aug, 2021.

[26] G. Oh, J. E. Lee, J. C. Ye, and J. C. Ye, "Unpaired MR Motion Artifact Deep Learning Using Outlier-Rejecting Bootstrap Aggregation," *IEEE Trans Med Imaging*, pp. 1-1, 2021.

[27] I. Oksuz, B. Ruijsink, E. Puyol-Anton, J. R. Clough, G. Cruz, A. Bustin, C. Prieto, R. Botnar, D. Rueckert, J. A. Schnabel, and A. P. King, "Automatic CNN-based detection of cardiac MR motion artefacts using k-space data augmentation and curriculum learning," *Medical Image Analysis,* vol. 55, pp. 136-147, Jul, 2019.

[28] C. Wang, Y. Liang, Y. Wu, S. Zhao, and Y. P. Du, "Correction of out-of-FOV motion artifacts using convolutional neural network," *Magn Reson Imaging,* vol. 71, pp. 93-102, Sep, 2020.

[29] Q. Zhang, E. Hann, K. Werys, C. Wu, I. Popescu, E. Lukaschuk, A. Barutcu, V. M. Ferreira, and S. K. Piechnik, "Deep learning with attention supervision for automated motion artefact detection in quality control of cardiac T1-mapping," *Artif Intell Med,* vol. 110, pp. 101955, Nov, 2020.

[30] O. Ronneberger, P. Fischer, and T. Brox, "U-Net: Convolutional Networks for Biomedical Image Segmentation," in 18th International Conference on Medical Image Computing and Computer-Assisted Intervention (MICCAI), Munich, Germany, 2015, pp. 234-241.

[31] S. Shah, P. Ghosh, L. S. Davis, and T. Goldstein, "Stacked U-Nets: a no-frills approach to natural image segmentation," *arXiv preprint arXiv:1804.10343*, 2018.

[32] D. Jha, M. A. Riegler, D. Johansen, P. Halvorsen, and H. D. Johansen, "DoubleU-Net: A Deep Convolutional Neural Network for Medical Image Segmentation," in IEEE 33rd International Symposium on Computer-Based Medical Systems (CBMS), Rochester, Minnesota, USA, 2020, pp. 558-564.

[33] S. Woo, J. Park, J.-Y. Lee, and I. S. Kweon, "CBAM: Convolutional Block Attention Module," in 15th Proceedings of the European Conference on Computer Vision (ECCV), Munich, Germany, 2018, pp. 3-19.

[34] S. R. Kecskemeti, and A. L. Alexander, "Test-retest of automated segmentation with different motion correction strategies: A comparison of prospective versus retrospective methods," *Neuroimage,* vol. 209, Apr 1, 2020.

[35] S. Lee, S. Jung, K.-J. Jung, and D.-H. Kim, "Deep Learning in MR Motion Correction: a Brief Review and a New Motion Simulation Tool (view2Dmotion)," *Investigative Magnetic Resonance Imaging,* vol. 24, no. 4, pp. 196-206, 2020.

[36] J.-Y. Zhu, T. Park, P. Isola, and A. A. Efros, "Unpaired image-to-image translation using cycle-consistent adversarial networks," in In Proceedings of the IEEE International Conference on Computer Vision (ICCV), Venice, Italy, 2017, pp. 2223-2232.